\documentclass[12pt]{article}
\usepackage{graphicx}
\usepackage{graphics}
\title{Series representations of the Riemann and Hurwitz zeta functions and series and integral representations of the first Stieltjes constant}
\author{Mark W. Coffey\\
Department of Physics\\
Colorado School of Mines\\
Golden, CO  80401\\
(Received $\mbox{~~~~~~~~~~~~~~~~~~~~~~~~~~~~~~~2011}$)}
\date{June 13, 2011}
\begin{document}
\maketitle
\baselineskip=25 pt
\begin{abstract}

We develop series representations for the Hurwitz and Riemann zeta functions in terms
of generalized Bernoulli numbers (N\"{o}rlund polynomials), that give the analytic
continuation of these functions to the entire complex plane.  Special cases yield 
series representations of a wide variety of special functions and numbers, including
log Gamma, the digamma, and polygamma functions.  A further byproduct is that $\zeta(n)$ values emerge as nonlinear Euler sums in terms of generalized harmonic numbers.  We additionally obtain series and integral representations of the first Stieltjes constant $\gamma_1(a)$. The presentation unifies some earlier results.

\end{abstract}
 
\baselineskip=15pt
\centerline{\bf Key words and phrases}
\medskip 

\noindent

Hurwitz zeta function, Riemann zeta function, Dirichlet $L$ function, generalized Bernoulli number, N\"{o}rlund polynomials and numbers, integral representation, series representation, digamma function, polygamma functions 

\vfill
\centerline{\bf 2010 AMS codes} 
11M06, 11Y60, 11M35

\baselineskip=25pt
\pagebreak
\medskip
\centerline{\bf Introduction and statement of results}
\medskip

Let $\zeta(s,a)$ be the Hurwitz zeta function and $\zeta(s)=\zeta(s,1)$ the Riemann zeta function \cite{ivicbk,riemann,titch}.    
We develop series representations for these functions in terms of coefficients with
the generalized Bernoulli numbers $B_n^{(\alpha)}$, and these have several consequences.
We illustrate that our results also extend to Dirichlet $L$ functions.  
As Corollaries, we obtain series representations of the log Gamma function, the
polygamma functions, and special numbers including the ordinary and generalized
harmonic numbers.  Specific application of our series is to the Stieltjes constants $\gamma_k(a)$, leading to new integral representations.  
We are able to unify some earlier results, specifically
including the very recent ones of Rubinstein \cite{rubin}.  The Discussion section
gives details as to how our framework subsumes that presentation.  
While an experimental mathematics approach lead to a representation and demonstration
of certain derivative values $\alpha_k'(1)$, this is a natural result
within our approach.  We then provide some other observations concerning the N\"{o}rlund numbers $B_n^{(n)}$, implications of our series representation of $\zeta(s,a)$, and finish with selected concluding remarks.


The generalized Bernoulli numbers are explicitly given by \cite{todorov}
$$B_n^{(\alpha)}=\sum_{k=0}^n {{\alpha+n} \choose {n-k}}{{\alpha+k-1} \choose k}
{{n!} \over {(n+k)!}}\sum_{j=0}^k (-1)^j {k \choose j} j^{n+k},  \eqno(1.1)$$
wherein $\alpha$ may be complex.
On the right side, a factor with the Stirling number of the second kind $S$ is evident,
$$k! S(n,k)=\sum_{j=0}^k (-1)^{k-j} {k \choose j} j^n.  \eqno(1.2)$$
$B_n^{(\alpha)}$ is a rational polynomial of degree $n$, with highest coefficient
$(-1/2)^n$.  The first few such are $B_0^{(\alpha)}=1$, $B_1^{(\alpha)}=-\alpha/2$, 
$B_2^{(\alpha)}=\alpha(3\alpha-1)/12$, and $B_3^{(\alpha)}=-\alpha^2(\alpha-1)/8$.
For these N\"{o}rlund polynomials, $\alpha=1$ is a simple root if $n>1$ is odd,
and $\alpha=0$ is a simple root if $n>0$ is even, and a double root if $n>1$ is odd.

The generalized Bernoulli polynomials (e.g., \cite{norlund,todorov}) are given by
$$B_n^{(\alpha)}(x)=\sum_{k=0}^n {n \choose k}B_k^{(\alpha)} x^{n-k},  \eqno(1.3)$$
$B_n^{(\alpha)}(x)=(-1)^n B_n^{(\alpha)}(\alpha-x)$, and they enter the generating function
$$\left({z \over {e^z-1}}\right)^\alpha e^{xz}=\sum_{n=0}^\infty B_n^{(\alpha)}(x)
{z^n \over {n!}}, ~~~~~~|z| < 2\pi.  \eqno(1.4)$$
As usual, $B_n^{(1)}(x)=B_n(x)$ and $B_n^{(1)}(0)=B_n(0)=B_n$ denote the Bernoulli
polynomials and numbers, respectively.

Throughout we write $\sigma=$ Re $s$.  
In the following $\Gamma$ is the Gamma function, $(a)_j=\Gamma(a+j)/\Gamma(a)$ is the Pochhammer symbol, $\psi=\Gamma'/\Gamma$ is the digamma function, $\gamma=-\psi(1)$ is the Euler constant, $\psi^{(j)}$ is the polygamma function, and $_pF_q$ is the generalized hypergeometric function (e.g., \cite{nbs,andrews,grad,watson}).

The Hurwitz zeta function, defined by $\zeta(s,a)=\sum_{n=0}^\infty (n+a)^{-s}$
for $\sigma >1$ and Re $a>0$ extends to a meromorphic function in the entire
complex $s$-plane.  In the Laurent expansion 
$$\zeta(s,a)={1 \over {s-1}}+ \sum_{n=0}^\infty {{(-1)^n} \over {n!}}\gamma_n(a)
(s-1)^n, \eqno(1.5)$$
$\gamma_n(a)$ are the Stieltjes constants \cite{coffeyjmaa,coffeystdiffs,coffey2009,stieltjes,wilton2},
and by convention one takes $\gamma_k = \gamma_k(1)$.

We have
{\newline \bf Proposition 1}.  Let $s\neq 1$ and Re $a>0$.  Then we have 
$$\zeta(s,a)={{\Gamma(a)} \over {\Gamma(s-1)}}\sum_{k=0}^\infty {{(-1)^k} \over {k!}}
{B_k^{(s+k-1)} \over {(s+k-1)}}{{\Gamma(s+k-1)} \over {\Gamma(s+k+a-1)}}.  \eqno(1.6)$$
In particular,
$$\zeta(s)={1 \over {\Gamma(s)}}{1 \over {(s-1)}}+{1 \over {\Gamma(s-1)}}\sum_{k=1}^ \infty {{(-1)^k} \over {k!}} {B_k^{(s+k-1)} \over {(s+k-1)^2}}.  \eqno(1.7)$$
Equivalently to (1.6) we have for Re $a>0$,
$$\Gamma(s)[s\zeta(s+1,a)-a^{-s}]=\Gamma(a)\sum_{k=0}^\infty {{(-1)^{k-1}} \over {k!}}
\left[{{(s-1)B_k^{(s+k-1)}} \over {s+k-1}}-{{sB_k^{(s+k)}} \over {s+k}}\right]{{\Gamma(
s+k)} \over {\Gamma(s+k+a)}}.  \eqno(1.8)$$

{\bf Corollary 1}.  For integers $n \geq 1$ we have
$$\psi^{(n)}(x)=(-1)^{n+1} n\Gamma(x)\sum_{k=0}^\infty {{(-1)^k} \over {k!}} {B_k^{(n+k)}
\over {(n+k)}} {{(n+k-1)!} \over {\Gamma(n+k+x)}}.  \eqno(1.9)$$
This representation holds due to the relation $\psi^{(n)}(x)=(-1)^{n+1}n! \zeta(n+1,x)$.
In particular, for the trigamma function we recover
$$\psi'(x)=\Gamma(x)\sum_{k=0}^\infty {{k!} \over {(k+1)}}{1 \over {\Gamma(x+k+1)}}
=\sum_{k=0}^\infty {{k!} \over {(k+1)}}{1 \over {(x+k)}}{1 \over {(x)_k}},  \eqno(1.10)$$
and the functional equation $\psi'(x+1)=\psi'(x)-1/x^2$:
$$\psi'(x+1)=\psi'(x)-\sum_{k=0}^\infty {{k!\Gamma(x)} \over {\Gamma(x+k+2)}}
=\psi'(x)-{{\Gamma^2(x)} \over {\Gamma^2(x+1)}}=\psi'(x)-{1 \over x^2}.  \eqno(1.11)$$
In the case of the tetragamma function, the functional equation follows from
$$\psi''(x+1)=\psi''(x)-2\Gamma(x)\sum_{k=1}^\infty {{k! H_k} \over {\Gamma(x+k+2)}}, \eqno(1.12)$$
where $H_k$ is the $k$th harmonic number, and the latter sum is given by
$$\sum_{k=1}^\infty {{k!} \over {\Gamma(x+k+2)}} \sum_{\ell=1}^k {1 \over \ell}
=\sum_{\ell=1}^\infty {1 \over \ell}\sum_{k=\ell}^\infty {{k!} \over {\Gamma(x+k+2)}}$$
$$={1 \over x}\sum_{\ell=1}^\infty {{\Gamma(\ell)} \over {\Gamma(x+\ell+1)}}={1 \over {x^3\Gamma(x)}}.  \eqno(1.13)$$
In connection with (1.10)-(1.13), some elementary relations have been relegated to
Appendix A.

We recover a well known value in the next result.  
{\newline \bf Corollary 2}.  We have
$$\zeta(0,a)={1 \over 2}-a=-B_1(a), \eqno(1.14)$$
where $B_1$ is the first Bernoulli polynomial.

We also have series representations for the digamma and log Gamma functions, and for
the first Stieltjes constant.
{\newline \bf Proposition 2}.  Let Re $a>0$.  Then (a)
$$\psi(a)-\ln a = \sum_{n=1}^\infty {{(-1)^n} \over n}{1 \over {(a)_n}}\left[B_n^{(n)}
+nB_{n-1}^{(n-1)}\right],  \eqno(1.15)$$
(b) for Re $x>-1$,
$$\ln \Gamma(x+1)=\sum_{k=0}^\infty {{(-1)^{k+1}} \over {k!}}B_k^{(k)}(1)\left[{1 \over
{k-1}}-{x \over k}-{{\Gamma(k-1)\Gamma(x+1)} \over {\Gamma(k+x)}}\right],  \eqno(1.16)$$
and (c)
$${1 \over 2}\ln^2 a+\gamma_1(a)=\sum_{n=1}^\infty {{(-1)^{n-1}} \over {n!}}[B_n^{(n)}
+nB_{n-1}^{(n-1)}]\sum_{k=0}^{n-1} (-1)^k {{n-1} \choose k}{{\ln(k+a)} \over {(k+a)}}.
\eqno(1.17)$$
We emphasize that the sum on the right side of (1.16) is well defined.  For $k$ near
$1$ we have $\Gamma(k-1)=1/(k-1)-\gamma+O(k-1)$ and for $k$ near $0$ we have
$\Gamma(k-1)=-1/k+\gamma-1+O(k)$.  Then the quantity on the right side in brackets is
$-x[\psi(x)+\gamma]+x-1$ for $k=0$ and $\psi(x+1)-x+\gamma$ for $k=1$.

{\bf Proposition 3}.  (a) For Re $a>0$ we have
$$\gamma_1(a)={1 \over 2}[\psi'(a)-\psi^2(a)]-{1 \over a}\int_0^1 v ~_4F_3(1,1,1,v+1;
2,2,a+1;1)dv, \eqno(1.18)$$
(b) we have
$$\gamma_1={\pi^2 \over 6}+\int_0^1 \left(\gamma \psi(x)+{1 \over 2}[\psi^2(x)-\psi'(x)]
\right)dx, \eqno(1.19)$$
and 
$$\gamma_1={\pi^2 \over 6}-{1 \over 2}\int_0^1\left[2{\gamma \over x}+\psi'(x)-\psi^2(x)
\right]dx,  \eqno(1.20)$$
and (c)
$$\gamma_2(a)={1 \over 3}\left[-\psi^3(a)+3\psi(a)\psi'(a)-\psi''(a)\right]
+2\Gamma(a)\sum_{k=1}^\infty {{(-1)^k} \over {k^2}}{1 \over {\Gamma(k+a)}}\left[
{{dB_k^{(s+k-1)}} \over {ds}}\right]_{s=1}$$
$$+2\Gamma(a) \sum_{k=1}^\infty {{(-1)^k} \over {k^3}}{B_k^{(k)} \over {\Gamma(k+a)}}[-1+\gamma k+k\psi(k)-k\psi(k+a)].  \eqno(1.21)$$ 

{\bf Corollary 3}.  We have
$$2[\zeta(2)-\gamma_1]=\gamma^2+1+2\sum_{k=2}^\infty (-1)^k {{\zeta(k)} \over {k-1}}
-\sum_{m=4}^\infty {{(-1)^m} \over {m-1}}\sum_{k=2}^{m-2} \zeta(k)\zeta(m-k).  \eqno(1.22)$$

A sum of (1.22) has many alternative forms, a few of which are collected in the
following.  We let $\Gamma(x,y)$ be the incomplete Gamma function.
{\newline \bf Corollary 4}.  
$$\sum_{k=1}^\infty {{(-1)^{k+1}} \over k}\zeta(k+1)=\int_0^\infty [\gamma+\Gamma(0,t)
+\ln t]{{dt} \over {e^t-1}}$$
$$=\sum_{j=1}^\infty {1 \over j}\ln\left(1+{1 \over j}\right)$$
$$=\int_0^\infty[\gamma+\psi\left(1+e^{-t}\right)]dt=\int_0^\infty[\gamma+\psi\left(
e^{-t}\right)+e^t]dt.  \eqno(1.23)$$
This sum, with approximate numerical value 1.25774688694, has been encountered before
in analytic number theory \cite{coffeysgamma} in the form $\sum_{k=1}^\infty \ln(k+1)/
[k(k+1)]$.  The latter reference gives several other representations of this sum.

A recursion for the derivatives of the N\"{o}rlund polynomials is given in the following.
{\newline \bf Proposition 4}.  For $n \geq 0$,
$$\partial_\alpha B_n^{(\alpha)}=-{n \over 2}B_{n-1}^{(\alpha)}-\sum_{j=0}^{n-2} {n \choose j} {B_{n-j} \over {(n-j)}}B_j^{(\alpha)}.  \eqno(1.24)$$
With a (null) convention for the sum, this includes the initial cases $\partial_\alpha B_0^{(\alpha)}=0$ and $\partial_\alpha B_1^{(\alpha)}=-1/2$.

Proposition 1 can be applied to Dirichlet $L$-functions.  As a first quick example, the Dirichlet $L$-function defined by
$$L(s) \equiv \sum_{n=0}^\infty {{(-1)^n} \over {(2n+1)^s}}, ~~~~~~~~~~~~
\mbox{Re} ~s>1, \eqno(1.25)$$
corresponding to quadratic characters modulo 4, can be expressed as 
$$L(s)=4^{-s}[\zeta(s,1/4)-\zeta(s,3/4)]=1+4^{-s}[\zeta(s,5/4)-\zeta(s,3/4)].
\eqno(1.26)$$
In particular, we have for nonnegative integers $m$ the special values 
$$L(2m+1)=-{{(2\pi)^{2m+1}} \over {2(2m+1)!}}B_{2m+1}(1/4).  \eqno(1.27)$$
Generally the values of $L$ at odd or even integer argument may be expressed in
terms of Euler or Bernoulli polynomials at rational argument and these in turn expressed in terms of the Hurwitz zeta function.  Therefore we may in this way
obtain many other computable series representations for $L(2m)$ and $L(2m+1)$.  
These include the special cases of $L(1)=\pi/4$, $L(2)=G \simeq 0.91596559$, Catalan's constant, and $L(3)=\pi^3/32$.  

In general, Dirichlet $L$ functions may be written as a combination of Hurwitz 
zeta functions.  For instance, for $\chi$ a principal (nonprincipal) character modulo $m$ and $\sigma > 1$ ($\sigma > 0$) we have
$$L(s,\chi) = \sum_{k=1}^\infty {{\chi(k)} \over k^s} ={1 \over m^s}\sum_{k=1}^m 
\chi(k) \zeta\left(s,{k \over m}\right).  \eqno(1.28)$$

\medskip
\centerline{\bf Proof of Propositions}

{\it Proposition 1}.  We will use the generating function \cite{norlund} (p. 147)
$$\left({{\ln(x+1)} \over x}\right)^z=z \sum_{k=0}^\infty {x^k \over {k!}}{B_k^{(z+k)}
\over {(z+k)}}, ~~~~~~|x| <1, \eqno(2.1)$$
valid for complex $z$.  We have for $\sigma>1$
$$\zeta(s,a)={1 \over {\Gamma(s)}}\int_0^\infty {{t^{s-1} e^{-(a-1)t}} \over {e^t-1}}
dt$$
$$={1 \over {\Gamma(s)}}\int_0^\infty {{t^{s-1} e^{-at}} \over {1-e^{-t}}} dt$$
$$={1 \over {\Gamma(s)}}\int_0^1 {{[-\ln(1-u)]^{s-1}} \over u} (1-u)^{a-1}du$$
$$={{(s-1)} \over {\Gamma(s)}}\sum_{k=0}^\infty {{(-1)^k} \over {k!}} {{B_k^{(s+k-1)}}
\over {(s+k-1)}} \int_0^1 u^{s+k-2}(1-u)^{a-1} du$$
$$={1 \over {\Gamma(s-1)}}\sum_{k=0}^\infty {{(-1)^k} \over {k!}} {{B_k^{(s+k-1)}}
\over {(s+k-1)}}B(s+k-1,a), \eqno(2.2)$$
where the Beta function factor
$$B(s+k-1,a)={{\Gamma(a)\Gamma(s+k-1)} \over {\Gamma(s+k+a-1)}}. \eqno(2.3)$$
With $B_k^{(s+k-1)}$ being polynomials in $s$, (2.2) gives the analytic continuation
of $\zeta(s,a)$ to the whole complex plane, completing the Proposition.

(1.8) is based upon the use of (2.1) and the integral
$$\int_0^1\left[{1 \over {\ln(1-y)}}+{1 \over y}\right](1-y)^{a-1} \ln^s\left({1 \over
{1-y}}\right)dy=\Gamma(s)[s\zeta(s+1,a)-a^{-s}].  \eqno(2.4)$$
Based upon (2.11) below for $a^{-s}\Gamma(s)$, we see the equivalence of (1.7)
and (1.5).  For (1.7) we may note the particular value $[B_k^{(k-1)}/(k-1)]_{k=1}=-
\int_0^1 tdt=-1/2$.

{\it Corollary 2}.  We recall that $B_0^{(\alpha)}=1$ and $B_1^{(s)}/s=-1/2$.
When $s \to 0$ in (1.6), only the $k=0$ and $k=1$ terms contribute, as
$1/\Gamma(s-1)=-s+(1-\gamma)s^2+O(s^3)$.  We obtain $\zeta(0,a)=1-a-1/2=1/2-a$.

{\it Remarks}.  With $B_n^{(\alpha)}(x)$ the generalized Bernoulli polynomial for $\nu \leq n$,
$$B_\nu^{(n+1)}(x)={{\nu!} \over {n!}}{d^{n-\nu} \over {dx^{n-\nu}}}(x-1)(x-2) \cdots
(x-n)=(-1)^n {{\nu!} \over {n!}}{d^{n-\nu} \over {dx^{n-\nu}}} (1-x)_n.  \eqno(2.5)$$
We have $B_n^{(\alpha)}=B_n^{(\alpha)}(0)$.  It follows as a very special case of (2.5)
that $B_k^{(k+1)}=(-1)^k k!$.  This latter relation allows the recovery of the series
$\zeta(2)=\sum_{k=1}^\infty 1/k^2$ from (1.7).

As also follows from (2.5), $B_k^{(k+2)}=(-1)^k k!H_{k+1}$, where $H_n=\sum_{k=1}^n 1/k$
is the $n$th harmonic number.  With these values, we have from (1.7) an Euler series
(e.g., \cite{coffey2003}, Appendix B) for $\zeta(3)$:  
$$\zeta(3)={1 \over 4}+\sum_{k=1}^\infty {H_{k+1} \over {(k+2)^2}}
=\sum_{k=1}^\infty {H_k \over {(k+1)^2}}.  \eqno(2.6)$$

Integral arguments of the digamma and polygamma functions are directly related
to the harmonic $H_n$ and generalized harmonic $H_n^{(r)}$ numbers:
$H_n = \psi(n+1)-\psi(1) = \psi(n+1)+\gamma$,   
$$H_n^{(r)}={{(-1)^{r-1}} \over {(r-1)!}}\left[\psi^{(r-1)}(n+1)-\psi^{(r-1)}
(1) \right ], \eqno(2.7)$$
Therefore Corollary 1 gives another form of these special numbers.  The
generalized harmonic numbers are given by
$$H_n^{(r)} \equiv \sum_{j=1}^n {1 \over j^r}, ~~~~H_n \equiv H_n^{(1)}. 
\eqno(2.8)$$
As also follows from (2.5), we have the relation
$$B_n^{(n+3)}=(-1)^n n![H_{n+2}^2-H_{n+2}^{(2)}].  \eqno(2.9)$$

Proceeding as for the Proposition, we have for Re $a>0$,
$$a^{-s}\Gamma(s)=\int_0^\infty x^{s-1}e^{-ax}dx=\int_0^1 [-\ln(1-t)]^{s-1}(1-t)^{a-1}dt, \eqno(2.10)$$
so that using (2.1) we have
$$a^{-s}\Gamma(s)=(s-1)\sum_{k=0}^\infty {{(-1)^k} \over {k!}} {B_k^{(s+k-1)} \over {(s+k-1)}}{{\Gamma(a)\Gamma(s+k)} \over {\Gamma(s+k+a)}}.  \eqno(2.11)$$

We have found that (1.9) and (1.15) are given in N\"{o}rlund's book \cite{norlund} in a section dealing with numerical differentiation and integration (pp. 243 and 244, respectively).

{\it Proposition 2}. (a) We introduce the constants (\cite{coffey2009}, Proposition 11)
$$p_{n+1}=-{1 \over {n!}}\int_0^1 (-x)_n dx={{(-1)^{n+1}} \over {n!}}\sum_{k=1}^n
{{s(k,n)} \over {k+1}}, \eqno(2.12)$$
where $s(k,\ell)$ is the Stirling number of the first kind.  These constants enter the generating function 
$$\sum_{n=1}^\infty p_{n+1}z^{n-1} ={1 \over z}+{1 \over {\ln(1-z)}}, ~~~~~~|z|<1.
\eqno(2.13)$$

{\bf Lemma 1}.  We have
$$k!p_{k+1}=(-1)^{k-1}[B_k^{(k)}+kB_{k-1}^{(k-1)}].  \eqno(2.14)$$

We have
$$\int_0^1 (x)_kdx=\int_0^1(1-y)_kdy=(-1)^k B_k^{(k)}=\int_0^1 (k-y)(1-y)_{k-1}dy$$
$$=-\int_0^1 y(1-y)_{k-1}dy+k\int_0^1 (1-y)_{k-1}dy, \eqno(2.15)$$
giving $(-1)^kB_k^{(k)}=-k!p_{k+1}+k(-1)^{k-1}B_{k-1}^{(k-1)}$, from which the Lemma
follows.

Part (a) of the Proposition then immediately follows from Proposition 5(a) of \cite{coffeyjnt}.  Here we give direct verifications:
$$\sum_{n=1}^\infty {{(-1)^n} \over n}{1 \over {(a)_n}}\left[B_n^{(n)} +nB_{n-1}^{(n-1)}
\right]=\sum_{n=1}^\infty {1 \over n}{1 \over {(a)_n}}\int_0^1[(x)_n-n(x)_{n-1}]dx$$
$$=\sum_{n=1}^\infty {1 \over n}{1 \over {(a)_n}}\int_0^1 (x-1)_n dx
=\int_0^1[\psi(a)-\psi(a-x+1)]dx=\psi(a)-\ln a.  \eqno(2.16)$$

Otherwise, we may employ
$${1 \over {(a)_n}}={{\Gamma(a)} \over {\Gamma(a+n)}}={1 \over {\Gamma(n)}}\int_0^\infty
e^{-at}(1-e^{-t})^{n-1} dt, ~~~~~~\mbox{Re} ~a>0, \eqno(2.17)$$
to write
$$\sum_{n=1}^\infty {{(-1)^n} \over n}{1 \over {(a)_n}}\left[B_n^{(n)}+nB_{n-1}^{(n-1)}
\right]=\int_0^1 \int_0^\infty e^{-at}{{(e^t-e^{xt})} \over {(1-e^t)}}dtdx$$
$$=\int_0^\infty e^{-at}\left[{1 \over {e^{-t}-1}}+{1 \over t}\right]dt
=\psi(a)-\ln a. \eqno(2.18)$$
The latter integral representation is a standard one (e.g., \cite{grad}, p. 943).

(b) From (2.1) and $B_k^{(n+1)}(1)=\left(1-{k \over n}\right)B_k^{(n)}$ it follows that
$$\left({x \over {\ln(x+1)}}\right)^z=\sum_{k=0}^\infty {x^k \over {k!}}B_k^{(k-z+1)}(1).
\eqno(2.19)$$
We use the integral representation \cite{jordan} (p. 343)
$$\ln \Gamma(x+1)=\int_0^1 {{[1-xt-(1-t)^x]} \over {t\ln(1-t)}}dt
=\sum_{k=0}^\infty {{(-1)^{k+1}} \over {k!}} B_k^{(k)}(1) \int_0^1 [1-xt-(1-t)^x]t^{k-2}
dt, \eqno(2.20)$$
with the last integral given by the Beta function $B(k-1,x+1)$, and part (b) follows.

For (c) we may use Lemma 1 and Proposition 5(b) of \cite{coffeyjnt}, or else apply
the integral representation (2.4). 

{\it Remarks}.  It is easy to show that
$$B_n^{(n)}=\sum_{k=0}^n (-1)^k {{s(n,k)} \over {k+1}}, \eqno(2.21)$$
where $s(n,n)=1$ and $s(n,0)=\delta_{n0}$ in terms of the Kronecker symbol $\delta_{jk}$.

From \cite{norlund} (p. 147), $(k-1)B_k^{(k)}(1)=-B_k^{(k-1)}=(k-1)(-1)^{k-1}
\int_0^1 t(1-t)_{k-1}dt$, so that for the coefficients of part (b) we have
$B_k^{(k)}(1)=(-1)^{k-1} \int_0^1 t(1-t)_{k-1}dt=(-1)^{k-1}k!p_{k+1}$.

In fact we have \cite{norlund} (p. 148)
$$(-1)^n (1-x)_n=\sum_{r=0}^n {n \choose r} x^r B_{n-r}^{(n+1)}.  \eqno(2.22)$$
Therefore, performing the integrations, we obtain
$$\int_0^1 (x)_ndx= (-1)^n \sum_{r=0}^n {n \choose r} {B_{n-r}^{(n+1)} \over {(r+1)}},
\eqno(2.23)$$
for $n>0$,
$$(-1)^n \int_0^1 x(1-x)_ndx= -{1 \over n}B_{n+1}^{(n)}
=(-1)^n \sum_{r=0}^n {n \choose r} {B_{n-r}^{(n+1)} \over {(r+2)}}, \eqno(2.24)$$
and
$$(-1)^n \int_0^1 (-x)_ndx= (-1)^n \sum_{r=0}^n {n \choose r} (2^{r+1}-1){B_{n-r}^{(n+1)} \over {(r+1)}}.  \eqno(2.25)$$

We record the latter result in the following form for the constants of (2.13).
\newline{\bf Lemma 2}.
$$p_{n+1}={{(-1)^{n+1}} \over {n!}} \sum_{r=0}^n {n \choose r} (2^{r+1}-1) {B_{n-r}^{(n+1)} \over {(r+1)}}.  \eqno(2.26)$$

From differentiating the generating function (1.4) there is the relation
$B_k^{(\alpha+1)}={{\alpha-k} \over \alpha}B_k^{(\alpha)}-kB_{k-1}^{(\alpha)}$.
At $\alpha=k-1$ this yields $k!p_{k+1}=(-1)^{k-1}B_k^{(k-1)}/(k-1)$.  According to
the relations above (2.22), this expression is equivalent to (2.12).

Comparing (2.1) with the generating function
$$[\ln(x+1)]^n=n! \sum_{k=n}^\infty s(k,n){x^k \over {k!}}, \eqno(2.27)$$
we have
$$s(m,n)={{(m-1)!} \over {(n-1)!}}{B_{m-n}^{(m)} \over {(m-n)!}}.  \eqno(2.28)$$
A generalization of Stirling numbers of the first kind in the first argument
is presented in Appendix B.

{\it Proposition 3}.  With $B_0^{(\alpha)}=1$, Proposition 1 gives
$$\zeta(s,a)={{\Gamma(a)} \over {\Gamma(s+a-1)}}{1 \over {(s-1)}} + 
{{\Gamma(a)} \over {\Gamma(s-1)}}\sum_{k=1}^\infty {{(-1)^k} \over {k!}}
{B_k^{(s+k-1)} \over {(s+k-1)}}{{\Gamma(s+k-1)} \over {\Gamma(s+k+a-1)}}.  \eqno(2.29)$$
We then expand both sides about $s=1$, using for instance
$${1 \over {\Gamma(s-1)}}=(s-1)+\gamma(s-1)^2+\left({\gamma^2 \over 2}-{\pi^2 \over
{12}}\right)(s-1)^3 + O[(s-1)^4], \eqno(2.30a)$$
and
$${{\Gamma(a)} \over {\Gamma(s+a-1)}}=1-\psi(a)(s-1)+{1 \over 2}[\psi^2(a)-\psi'(a)]
(s-1)^2$$
$$+ {1 \over 6}\left[-\psi^3(a)+3\psi(a)\psi'(a)-\psi''(a)\right](s-1)^3 + O[(s-1)^4].  \eqno(2.30b)$$
From the $O(s-1)$ terms we recover the known relation $\gamma_0(a)=-\psi(a)$.
In addition we have
$$\left[\zeta'(s,a)+{1 \over {(s-1)^2}}\right]_{s \to 1^+}=-\gamma_1(a)
={1 \over 2}[\psi^2(a)-\psi'(a)]+\Gamma(a)\sum_{k=1}^\infty {{(-1)^k} \over k^2}
{B_k^{(k)} \over {\Gamma(k+a)}}.  \eqno(2.31)$$
Now \cite{norlund} (p. 147)
$$B_k^{(k)}=(-1)^k \int_0^1 (1-t)_k dt=(-1)^k \int_0^1 (t)_k dt.  \eqno(2.32)$$
We easily have
$$\sum_{k=1}^\infty {1 \over k^2}{{(t)_k} \over {\Gamma(k+a)}}={1 \over {\Gamma(a+1)}}
t ~_4F_3(1,1,1,t+1;2,2,a+1;1), \eqno(2.33)$$
and part (a) readily follows.

Part (b) then follows at $a=1$.  
For (1.20), in light of $\psi(x)=-1/x -\gamma +(\pi^2/6)
x+O(x^2)$ as $x \to 0$, we write the term $\gamma \psi(x)=\gamma[\psi(x)+1/x-1/x]$.
We then employ the integral
$$\int_0^1 \left[\psi(x)+{1 \over x}\right]dx=\int_0^1 \psi(x+1)dx= \ln\Gamma(x+1)|_0^1
=0.  \eqno(2.34)$$

Part (c) proceeds similarly to (a), also using
$$B_k^{(s+k-1)}=B_k^{(k)}+\left[{{dB_k^{(s+k-1)}} \over {ds}}\right]_{s=1}(s-1)+
{1 \over 2}\left[{{d^2B_k^{(s+k-1)}} \over {ds^2}}\right]_{s=1}(s-1)^2+O[(s-1)^3].
\eqno(2.35)$$  
Expression (1.21) follows from
$$\left[\zeta''(s,a)-{2 \over {(s-1)^3}}\right]_{s \to 1^+}=\gamma_2(a).  \eqno(2.36)$$

{\it Remark}.  We have the Laurent expansion about $s=1$ of the logarithmic derivative of the zeta function, 
$${{\zeta'(s)} \over {\zeta(s)}}=-{1 \over {s-1}}-\sum_{p=0}^\infty 
\eta_p (s-1)^p, ~~~~~~|s-1| < 3, \eqno(2.37)$$ 
where $\eta_0=-\gamma$.  From Proposition 2 or Corollary 3 we then obtain corollary expressions for $\eta_1=\gamma^2+2\gamma_1$.

{\it Corollary 3}.  We use Proposition 3(b) along with the expansions (e.g., \cite{grad},
p. 944 or \cite{nbs}, p. 259)
$$\psi(x+1)=\psi(x)+{1 \over x}=-\gamma+\sum_{k=2}^\infty (-1)^k \zeta(k) x^{k-1}, 
~~~~~~|x| <1, \eqno(2.38)$$
$$\psi'(x)={1 \over x^2}+\sum_{k=1}^\infty (-1)^{k+1}k \zeta(k+1)x^{k-1}, \eqno(2.39)$$
and
$$\psi^2(x)={1 \over x^2}+2{\gamma \over x}+\gamma^2-2\sum_{k=2}^\infty (-1)^k \zeta(k)
x^{k-2}-2\gamma \sum_{k=2}^\infty (-1)^k\zeta(k)x^{k-1}$$
$$+\sum_{m=4}^\infty \sum_{k=2}^{m-2}(-1)^m \zeta(k)\zeta(m-k)x^{m-2}.  \eqno(2.40)$$
We insert these into 
$$2[\zeta(2)-\gamma_1]=\int_0^1\left[2{\gamma\over x}+\psi'(x)-\psi^2(x)\right]dx,
\eqno(2.41)$$
integrate term by term, Abel sum the series
$$\lim_{x \to 1^-} \sum_{k=1}^\infty (-1)^{k+1} \zeta(k+1)x^k=1, \eqno(2.42)$$
and use the easily proved sum 
$\sum_{k=2}^\infty (-1)^k \zeta(k)/k=\gamma$, giving the Corollary.

{\it Corollary 4}.  The first expression follows by using a standard integral
representation for the zeta function.  The next follows by use of a geometric series,
and the integral representation with the digamma function follows by using the Laplace
transform representation of $1/k$.

{\it Proposition 4}.  From the generating function (1.4) we have
$$\left({z \over {e^z-1}}\right)^\alpha e^{xz}\ln\left({z \over {e^z-1}}\right)=\sum_{n=1}^\infty {{\partial B_n^{(\alpha)}(x)} \over {\partial \alpha}}
{z^n \over {n!}}, ~~~~~~|z| < 2\pi.  \eqno(2.43)$$
In order to expand the log factor, we first note that
$${d \over {dz}}\ln\left({z \over {e^z-1}}\right)={1 \over z}-1-{1 \over {e^z-1}}
={1 \over z}-1-{1 \over z}\sum_{n=0}^\infty B_n {z^n \over {n!}}, \eqno(2.44)$$
yielding
$${d \over {dz}}\ln\left({z \over {e^z-1}}\right)=-1-\sum_{n=0}^\infty {B_{n+1} \over
{(n+1)!}}z^n.  \eqno(2.45)$$
Upon integrating,
$$\ln\left({z \over {e^z-1}}\right)=-z-\sum_{n=0}^\infty {{B_{n+1}z^{n+1}} \over
{(n+1)(n+1)!}}=-{z\over 2}-\sum_{n=1}^\infty {{B_{n+1}z^{n+1}} \over {(n+1)(n+1)!}}.
\eqno(2.46)$$
Then at $x=0$ in (2.43) we have
$$\sum_{n=1}^\infty {{\partial B_n^{(\alpha)}} \over {\partial \alpha}} {z^n \over {n!}}
=\sum_{n=0}^\infty B_n^{(\alpha)}(x){z^n \over {n!}}\left(-{z\over 2}-\sum_{m=1}^\infty {{B_{m+1}z^{m+1}} \over {(m+1)(m+1)!}}\right)$$
$$=-{1 \over 2}\sum_{n=0}^\infty B_n^{(\alpha)} {z^{n+1} \over {n!}}-\sum_{n=0}^\infty
\sum_{m=1}^\infty {{B_n^{(\alpha)}B_{m+1} z^{n+m+1}} \over {n!(m+1)!(m+1)}}.  \eqno(2.47)$$
Reordering the double sum,
$$\sum_{n=1}^\infty {{\partial B_n^{(\alpha)}} \over {\partial \alpha}} {z^n \over {n!}}
=-{1 \over 2}\sum_{n=0}^\infty B_n^{(\alpha)} {z^{n+1} \over {n!}}-\sum_{\ell=1}^\infty
\sum_{n=0}^{\ell-1} {{B_n^{(\alpha)}B_{\ell-n+1}} \over {n!(\ell-n+1)!}}{z^{\ell+1} \over
{(\ell-n+1)}}, \eqno(2.48)$$
from which the Proposition follows.

{\it Remarks}.  It is seen that the recursion (1.24) also applies for $x \neq 0$.
Since $B_{2n-1}=0$ for $n>1$, the sum of (2.44) may be written as
${1 \over 2}\sum_{m=1}^\infty {{B_{2m}z^{2m}} \over {m(2m)!}}$.
From (2.43) we have similarly for higher order derivatives
$$\left({z \over {e^z-1}}\right)^\alpha e^{xz}\ln^j\left({z \over {e^z-1}}\right)=\sum_{n=j}^\infty {{\partial^j B_n^{(\alpha)}(x)} \over {\partial \alpha^j}} {z^n \over {n!}}, ~~~~~~|z| < 2\pi,  \eqno(2.49)$$
implying
$$\sum_{n=j}^\infty \left({{\partial^j B_n^{(\alpha)}(x)} \over {\partial \alpha^j}} \right)_{\alpha=0} {z^n \over {n!}}=e^{xz}\ln^j\left({z \over {e^z-1}}\right).  \eqno(2.50)$$

\medskip
\centerline{\bf Discussion}  

Rubinstein \cite{rubin} developed several expansions for the Riemann zeta function
using certain polynomials $\alpha_k(s)$.  
That work employed the generating function
$$\left(-{{\ln(1-t)} \over t}\right)^{s-1}=\sum_{k=0}^\infty \alpha_k(s)t^k, ~~~~~~
|t|<1.  \eqno(3.1)$$
Comparing with (2.1) we identify
$$\alpha_k(s)=(s-1){{(-1)^k} \over {k!}}{B_k^{(s+k-1)} \over {(s+k-1)}}.  \eqno(3.2)$$
We have
$$\alpha_k'(s)={{(-1)^k} \over {(k-1)!}}{B_k^{(s+k-1)} \over {(s+k-1)^2}}+{{(-1)^k} \over {k!}}{{(s-1)} \over {(s+k-1)}}{d \over {ds}} B_k^{(s+k-1)}, \eqno(3.3)$$
where, by (1.1), and using the functional equation of the digamma function,
$$\partial_\alpha B_n^{(\alpha)}=[\psi(n+\alpha+1)-\psi(\alpha)]B_n^{(\alpha)}$$
$$-\sum_{k=0}^n {{\alpha+n} \choose {n-k}}{{\alpha+k-1} \choose k} {1 \over {(k+\alpha)}}{{n!} \over {(n+k)!}}\sum_{j=0}^k (-1)^j {k \choose j} j^{n+k},  \eqno(3.4)$$
and $\psi(n+\alpha+1)-\psi(\alpha)=\sum_{r=0}^n 1/(\alpha+r)$.
We have precisely by (2.32)
$$\alpha_k'(1)={{(-1)^k} \over {k!}}{B_k^{(k)} \over k}={1 \over {kk!}}\int_0^1 (t)_k dt,
\eqno(3.5)$$
that was found by indirect means in \cite{rubin}.

We also have for $k\geq 0$ and $m \geq 0$,
$$\alpha_k(-k)={{(-1)^k} \over {k!}}, \eqno(3.6a)$$
$$\alpha_{2m+2}(-2m-1)={B_{2m+2} \over {(2m+2)!}}, \eqno(3.6b)$$
and
$$\alpha_{2m+2}(-2m)=-(2m+1){B_{2m+2} \over {(2m+2)!}}. \eqno(3.6c)$$
By using the generating function (1.5), one easily sees that $B_n^{(-1)}=1/(n+1)$,
so that from (3.2), $\alpha_k(-k)=(k+1)(-1)^k B_k^{(-1)}/k!$, giving (3.6a).

\medskip
\centerline{\bf On the N\"{o}rlund numbers $B_n^{(n)}$}  

These numbers have the known asymptotic form $B_n^{(n)} \sim (-1)^n n!/\ln n$ as $n 
\to \infty$.
From (2.32) it is easy to see how a standard generating function for them arises:
$$\sum_{n=0}^\infty {{B_n^{(n)}} \over {n!}}z^n=\int_0^1 (1+z)^{-t}dt={z \over {(1+z)\ln(1+z)}}.  \eqno(4.1)$$
From (2.32) and the relation above (2.22) for $B_r^{(r-1)}$ it is also clear how 
to develop expressions given in \cite{norlund} (p. 244):
$$\ln\left({{x+1} \over x}\right)= \sum_{r=0}^\infty {{(-1)^r B_r^{(r)}} \over {(x+1)_r}},
\eqno(4.2)$$
and
$$\ln\left({{x+1} \over x}\right)= {1 \over x}-{1 \over {2x(x+1)}}-{1 \over x} \sum_{r=2}^\infty {{(-1)^r B_r^{(r-1)}} \over {(r-1)}}{1 \over {(x+1)_r}}.  \eqno(4.3)$$
We note that differentiation with respect to $x$ of these formulas gives further sum
identities for generalized Bernoulli numbers.

Given the generating function (2.19) and that for harmonic numbers,
$$\sum_{k=0}^\infty H_{k+1}z^k=-{{\ln(1-z)} \over {z(1-z)}}, \eqno(4.4)$$
one suspects various relations between generalized Bernoulli numbers and harmonic
numbers.  As an example, we present the following.  Although this relation could be
proven with generating functions, we give a proof employing special function theory.

{\bf Proposition 5}.  For integers $n \geq 0$,  
$$\sum_{r=0}^n {{(-1)^{n-r}} \over {(n-r)!}}B_{n-r}^{(n-r)} H_{r+1}=n+1.  \eqno(4.5)$$

{\it Proof}.  We begin by reordering sums and applying Chu-Vandermonde summation,
$$\sum_{r=0}^n {{(-1)^{n-r}} \over {(n-r)!}}B_{n-r}^{(n-r)} H_{r+1}= 
\sum_{r=0}^n {{(-1)^{n-r}} \over {(n-r)!}}B_{n-r}^{(n-r)} \sum_{\ell=1}^{r+1} {1 \over \ell}
=\sum_{\ell=1}^{n+1}{1 \over \ell}\sum_{r=\ell-1}^n {{(-1)^{n-r}} \over {(n-r)!}}B_{n-r}^{(n-r)}$$
$$=\sum_{\ell=1}^{n+1}{1 \over \ell}\sum_{r=\ell-1}^n {1 \over {(n-r)!}}\int_0^1 (t)_{n-r}dt
$$
$$=\sum_{\ell=1}^{n+1}{1 \over \ell}\int_0^1 {{\Gamma(t+n-\ell+2)} \over {\Gamma(t+1)
\Gamma(n-\ell+2)}}dt$$
$$=\int_0^1 {{\Gamma(t+n+2)} \over {\Gamma(n+2)\Gamma(t+1)}}[\psi(t+n+2)-\psi(t+1)]dt$$
$$={1 \over {n+1}}\int_0^1 \left({d \over {dt}}{1 \over {B(n+1,t+1)}}\right) dt=n+1.
\eqno(4.6)$$
Above, we manipulated a terminating $_3F_2$ function at unit argument, since
$$\sum_{\ell=1}^{n+1}{1 \over \ell}{{\Gamma(t+n-\ell+2)} \over {\Gamma(n-\ell+2)}}z^\ell=\sum_{\ell=0}^n {{(1)_\ell^2} \over {(2)_\ell}}
{{\Gamma(t+n-\ell+1)} \over {\Gamma(n-\ell+1)}} {z^{\ell+1} \over {\ell!}}$$
$$=\sum_{\ell=0}^n {{(1)_\ell^2} \over {(2)_\ell}}{{\Gamma(t+n+1)} \over
{\Gamma(n+1)}}{{(-n)_\ell} \over {(-t-n)_\ell}} {z^{\ell+1} \over {\ell!}}
=z {{\Gamma(t+n+1)} \over {\Gamma(n+1)}} ~_3F_2(1,1,-n;2,-t-n;z).  \eqno(4.7)$$
Because
$$_3F_2(1,1,a;2,b;1)=\left({{b-1} \over {a-1}}\right)[\psi(b-1)-\psi(b-a)],
\eqno(4.8)$$
we have
$${{\Gamma(t+n+1)} \over {\Gamma(n+1)}} ~_3F_2(1,1,-n;2,-t-n;1)={{\Gamma(t+n+1)} \over {\Gamma(n+1)}}{{(t+n+1)} \over {(n+1)}}[\psi(-t-n-1)-\psi(-t)]$$
$$={{\Gamma(t+n+2)} \over {\Gamma(n+2)}}[\psi(t+n+2)-\psi(t+1)], \eqno(4.9)$$
by recalling that $\psi(1-x)-\psi(x)=\pi \cot(\pi x)$.

\medskip
\centerline{\bf Other relations from (1.6)}  

From the representation (1.6) we have
$$\partial_a \zeta(s,a)=-s\zeta(s+1,a)=\psi(a) \zeta(s,a)$$
$$-{{\Gamma(a)} \over {\Gamma(s-1)}}\sum_{k=0}^\infty {{(-1)^k} \over {k!}}
{B_k^{(s+k-1)} \over {(s+k-1)}}{{\Gamma(s+k-1)} \over {\Gamma(s+k+a-1)}}\psi(s+k+a-1).  \eqno(5.1)$$
We also have the special case
$$\zeta\left(s,{1 \over 2}\right)={{\Gamma(1/2)} \over {\Gamma(s-1)}}\sum_{k=0}^\infty {{(-1)^k} \over {k!}}{B_k^{(s+k-1)} \over {(s+k-1)}}{{\Gamma(s+k-1)} \over {\Gamma(s+k-1/2)}}=(2^s-1)\zeta(s).  \eqno(5.2)$$
It would be desirable to otherwise have proofs of the right-most equalities in (4.1) and
(5.2).  The duplication formula for the Gamma function may be useful in this regard for
(5.2).

The Bernoulli polynomials satisfy the multiplication formula
$$B_n(mx)=m^{n-1}\sum_{k=0}^{m-1} B_n\left(x+{k \over m}\right),  \eqno(5.3)$$
that is easily verified with the generating function (1.4) with $\alpha=1$.
One may ask whether there is a generalization to a formula such as
$$B_n^{(\alpha)}(mx)=m^{n-\alpha}f_1(n,\alpha)\sum_{k=0}^{m-1} \left[B_n^{(\alpha)} \left(x+{{\alpha k} \over m}\right)+f_2(x,n,\alpha)\right],  \eqno(5.4)$$
where $f_1$ and $f_2$ are such that $f_1=1$ and $f_2=0$ for $\alpha=1$.

\medskip
\centerline{\bf Summary remarks}  

The series (1.6) and (1.7) converge everywhere in the complex plane and include as special cases the relations $\zeta(1-m,a)=-B_m(a)/m$ and the trivial zeros $\zeta(-2n)=0$. The treatment has been sufficiently general to subsume expansion of the zeta function at integer argument in terms of Stirling numbers.  We may emphasize, among other features, that nonlinear Euler sums for $\zeta(n)$ naturally emerge as special cases.  As an example, we have from (2.9)
$$\zeta(4)={1 \over 2}\sum_{k=2}^\infty {{[H_k^2-H_k^{(2)}]} \over {(k+1)^2}},  \eqno(6.1)$$
in terms of generalized harmonic numbers $H_n^{(r)}$.

Proposition 3 may be the only known integral representation for the first Stieltjes 
constant $\gamma_1$ in terms of the digamma and trigamma functions.

It seems that the second line of (3.4) for the derivatives $\partial_\alpha B_n^{(\alpha)}$ could have an alternative, more compact form, possibly involving
$B_j^{(\alpha)}$ and $B_n$ values.  Such an expression would be very convenient in
further developments for the Stieltjes and other constants.  For instance, it could be
immediately applied to Proposition 3(c).

As mentioned in connection with (5.4), further investigation of the N\"{o}rlund
polynomials appears to be in order.



\medskip
\centerline{\bf Appendix A:  Beta function-based relations}
\medskip  

We have for Re $x>0$ and $n \geq 0$ an integer, the special case of the Beta 
function
$$B(x,n+1)=\int_0^1 t^{x-1}(1-t)^ndt={{n!}\over {x(x+1)\cdots (x+n)}}={{n!} \over
{(x)_{n+1}}},  \eqno(A.1)$$
that may be verified by induction.  So as relates to (1.10)-(1.12), we have
$$\sum_{k=0}^\infty {{k!} \over {(k+1)}}{1 \over {(x)_{k+1}}}=\sum_{k=0}^\infty
{1 \over {(k+1)}}\int_0^1 t^{x-1}(1-t)^k dt$$
$$=\int_0^1 {t^{x-1} \over {t-1}}\ln t ~dt=\psi'(x).  \eqno(A.2)$$
Here, we have recalled (e.g. by \cite{grad}, p. 943)
$$\psi^{(j)}(z)=\int_0^1 {t^{z-1} \over {t-1}}\ln^j t ~dt, ~~~~~~\mbox{Re} ~z>0.
\eqno(A.3)$$

In regard to (1.12), we recall the generating function with harmonic numbers
$$\sum_{k=1}^\infty H_kz^k={{\ln(1-z)} \over {z-1}}={{\mbox{Li}_1(z)} \over {1-z}},
\eqno(A.4)$$
where Li$_j$ is the polylogarithm function, and $H_0=0$, and its integrated form,
$$\sum_{k=1}^\infty {H_k \over {k+1}}z^{k+1}={1 \over 2}\ln^2(1-z).  \eqno(A.5)$$
Then from (A.1) we have
$$\sum_{k=1}^\infty {{k! H_k} \over {(x)_{k+2}}}=\sum_{k=1}^\infty {H_k \over {k+1}}
\int_0^1 t^{x-1} (1-t)^{k+1} dt$$
$$={1 \over 2}\int_0^1 t^{x-1}\ln^2 t ~dt={1 \over 2}\int_0^\infty v^2 e^{-xv} dv
={1 \over x^3}.  \eqno(A.6)$$

By binomially expanding the integrand of (A.1) and/or by differentiating with respect
to $x$ we obtain other relations.  For instance, we have
$$\partial_x B(x,n+1)=\int_0^1 t^{x-1}(1-t)^n \ln t ~dt={{n!} \over {(x)_{n+1}}}
[\psi(x)-\psi(x+n+1)]=-\sum_{\ell=0}^n {{(-1)^\ell} \over {(x+\ell)^2}} {n \choose \ell}.
\eqno(A.7)$$

Alternative points of view of $B(x,n+1)=\sum_{\ell=0}^n {{(-1)^\ell} \over {(x+\ell)}}
{n \choose \ell}$ are in terms of partial fractions or as divided differences of $1/x$.

\medskip
\centerline{\bf Appendix B:  Generalized Stirling numbers of the first kind}
\medskip

We let $s(j,k)$ denote the Stirling numbers of the first kind with integer
arguments.  We proceed to generalize these with the first argument complex.
In the following $\psi$ again denotes the digamma function, $\psi^{(j)}$ the polygamma functions, and $(z)_k=\Gamma(z+k)/\Gamma(z)$ the Pochhammer symbol.

It is shown in \cite{coffey2009} (Lemma 1) that
$$\left.\left({d \over {ds}}\right)^\ell (s)_j \right|_{s=1}=(-1)^{j+\ell} \ell! s(j+1,\ell+1).  \eqno(B.1)$$
From this we make the extension to $\lambda \in C$
$$s(\lambda,k)={{(-1)^{\lambda +k}} \over {(k-1)!}}\left({d \over {ds}}\right)^{k-1}
\left. (s)_{\lambda-1}\right|_{s=1}.  \eqno(B.2)$$
We then obviously have agreement with the usual Stirling numbers of the first kind
when $\lambda$ is a nonnegative integer.  When $\lambda$ is a positive integer,
$(s)_{\lambda-1}$ is a polynomial of degree $\lambda-1$ in $s$.  Hence by the 
definition (B.2), $s(\lambda,k)=0$ when $k>\lambda$.  Moreover, we show that these extended Stirling numbers satisfy the same recursion relation as their classical counterparts.

{\bf Lemma B1}.  We have
$$s(\lambda,k)=s(\lambda-1,k-1)-(\lambda-1)s(\lambda-1,k).  \eqno(B.3)$$

{\it Proof}.  We first note $(s)_{\lambda-1}=(s+\lambda-2)(s)_{\lambda-2}$ so that from (B.2)
$$s(\lambda,k)={{(-1)^{\lambda +k}} \over {(k-1)!}}\left({d \over {ds}}\right)^{k-2}
\left[(s)_{\lambda-2} +(s+\lambda-2) {d \over {ds}}(s)_{\lambda-2}\right]_{s=1}$$
$$={{(-1)^{\lambda +k}} \over {(k-1)!}}\left[(-1)^{\lambda+k} (k-2)!s(\lambda-1,k-1)
+ \left({d \over {ds}}\right)^{k-2} (s+\lambda-2) {d \over {ds}}(s)_{\lambda-2} \right]_{s=1}.  \eqno(B.4)$$
By using the product rule we find for the last term
$$\left.\left({d \over {ds}}\right)^{k-2} (s+\lambda-2) {d \over {ds}}(s)_{\lambda-2} \right|_{s=1}=D_s^{k-2} (s)_{\lambda-2}|_{s=1}+(\lambda-1)D_s^{k-1}(s)_{\lambda-2} |_{s=1}$$
$$=(-1)^{\lambda+k}(k-2)!{{k-2} \choose {k-3}} s(\lambda-1,k-1)+(-1)^{\lambda+k-1}
(\lambda-1)(k-1)!s(\lambda-1,k).  \eqno(B.5)$$
Then (B.4) becomes
$$s(\lambda,k)={{(-1)^{\lambda +k}} \over {(k-1)!}}[(-1)^{\lambda+k}(k-1)(k-2)!
s(\lambda-1,k-1)+(-1)^{\lambda+k-1}(\lambda-1)(k-1)!s(\lambda-1,k)], \eqno(B.6)$$
and the Lemma follows.

{\it Remarks}.  We have
$${d \over {ds}}(s)_\lambda=(s)_\lambda[\psi(s+\lambda)-\psi(s)],  \eqno(B.7)$$
so that the `higher' $s(\lambda,k)$ values may be obtained via Bell polynomials.
We recall the harmonic $H_n$ and generalized harmonic $H_n^{(r)}$ numbers:
$H_n = \psi(n+1)-\psi(1) = \psi(n+1)+\gamma$,   
$$H_n^{(r)}={{(-1)^{r-1}} \over {(r-1)!}}\left[\psi^{(r-1)}(n+1)-\psi^{(r-1)}
(1) \right ]. \eqno(B.8)$$
Our definition of $s(\lambda,k)$ then extends as desired.  For instance, we have
$$s(\lambda,2)=(-1)^\lambda (\lambda-1)!H_{\lambda-1}, ~~~~~s(\lambda,3)=(-1)^{\lambda+1}
{{(\lambda-1)!} \over 2}[H_{\lambda-1}^2-H_{\lambda-1}^{(2)}],  \eqno(B.9a)$$
and
$$s(\lambda,3)=(-1)^\lambda {{(\lambda-1)!} \over 6}[H_{\lambda-1}^3-3H_{\lambda-1}
H_{\lambda-1}^{(2)}+2H_{\lambda-1}^{(3)}].  \eqno(B.9b)$$
Here, it is understood that $(\lambda-1)!=\Gamma(\lambda)$.

\pagebreak

\end{document}